\documentstyle[12pt,titlepage,epsfig]{article}

\font\medio=cmr10 scaled \magstep2
\outer\def\beginsection#1\par{\medbreak\bigskip
      \message{#1}\leftline{\bf#1}\nobreak\medskip
\vskip-\parskip
      \noindent}

\def\laq{\raise 0.4ex\hbox{$<$}\kern -0.8em\lower 0.62
ex\hbox{$\sim$}}
\def\gaq{\raise 0.4ex\hbox{$>$}\kern -0.7em\lower 0.62
ex\hbox{$\sim$}}

\def\beq{\begin{equation}}
\def\eeq{\end{equation}}
\def\bea{\begin{eqnarray}}
\def\eea{\end{eqnarray}}

\def \pa {\partial}
\def \ra {\rightarrow}

\def \fb {\overline \phi}
\def \fbp {\dot{\fb}}
\def \bp {\dot{\beta}}

\def \la {\lambda}
\def \ls {\lambda_s}
\def \La {\Lambda}

\def \b {\beta}

\def \ap {\alpha^{\prime}}

\def \Ga {\Gamma}

\def \da {\delta}

\def \Om {\Omega}

\def \pfb {\Pi_{\fb}}

\def \pbe {\Pi_{\b}}

\begin{document}
\bibliographystyle {unsrt}

\titlepage
\begin{flushright}
CERN-TH/96-322 \\
IFUP-TH/96-65 \\
hep-th/9701146\\
\end{flushright}
\vspace{6mm}
\begin{center}
{\bf EXPANDING AND CONTRACTING UNIVERSES\\
IN THIRD QUANTIZED STRING COSMOLOGY}\\

\vspace{10mm}

A. Buonanno${}^{(a) (b)}$, M. Gasperini${}^{(c) (d)}$,
M. Maggiore${}^{(a) (b)}$ and C. Ungarelli${}^{(a) (b)}$\\
\vspace{6mm}

${}^{(a)}$
{\sl  Dipartimento di Fisica, Universit\`a di Pisa, \\
Piazza Torricelli 2, I-56100 Pisa, Italy} \\
${}^{(b)}$
{\sl Istituto Nazionale di Fisica Nucleare, Sezione di Pisa, Pisa, Italy} \\
${}^{(c)}$
{\sl Theory Division, CERN, CH-1211 Geneva 23, Switzerland} \\
${}^{(d)}$
{\sl Dipartimento di Fisica Teorica, Universit\`a di Torino, \\
Via P. Giuria 1, 10125 Turin, Italy }\\
\end{center}
\vskip 1cm
\centerline{\medio  Abstract}

\noindent
We discuss the possibility of  quantum transitions from the string 
perturbative vacuum to cosmological configurations 
characterized by isotropic contraction and decreasing 
dilaton. When the dilaton potential preserves the sign of the Hubble 
factor throughout the evolution, such transitions 
can be represented as an anti-tunnelling of the 
Wheeler--De Witt wave function in minisuperspace or, in a 
third-quantization language, as the production of pairs of universes 
out of the vacuum.

\vspace{7mm}
\centerline{\sl To appear in {\bf Class. Quantum Grav.}}
\vspace{5mm}
\vfill
\begin{flushleft}
CERN-TH/96-322 \\
November 1996 
\end{flushleft}

\newpage

At very early times, according to the standard cosmological scenario,
the Universe is expected to approach a Planckian, quantum gravity
regime where a classical description of the spacetime manifold is no
longer appropriate. A possible quantum description of the Universe, in
that regime, is based on the Wheeler--De Witt (WDW) wave function
\cite{1,2}, generally defined on the superspace spanned by all
three-dimensional geometric configurations. In that context it becomes
possible to compute, with an appropriate model of (mini)superspace, the
probability distribution of a given cosmological configuration versus an 
appropriate ``state" parameter (for instance the cosmological constant 
$\La$). The results, however, are in general affected by 
operator-ordering ambiguities, and are also strongly dependent on the 
boundary conditions \cite{3}--\cite{5} 
imposed on the solutions of the WDW equation.

String theory has recently motivated the study of a cosmological
scenario in which the Universe starts from the string perturbative
vacuum  and evolves through an initial, ``pre-big bang" phase \cite{6}, 
characterized by an accelerated growth of the curvature and of the
gauge coupling $g=e^{\phi/2}$ ($\phi$ is the dilaton field).  
In such a context, the WDW equation
is obtained from the low-energy string effective 
action \cite{7}--\cite{9}, and has no
operator ordering ambiguities \cite{7} 
since the ordering is uniquely fixed by the
duality symmetries of the action. Also, the boundary conditions are 
determined by the choice of the perturbative vacuum as the initial 
state for the cosmological evolution.

According to the lowest-order effective action, the classical evolution 
from the perturbative vacuum necessarily leads the background 
to a singularity, and 
the transition to the present decelerated ``post-big bang" 
configuration is impossible, for any realistic type of (local) dilaton 
potential \cite{10}. With an appropriate potential, however, the 
transition may become allowed at the quantum level even if, for the same 
potential, it remains classically forbidden. 
This effect was discussed in previous papers \cite{7}, in which the WDW 
equation was applied to compute the transition probability between 
two duality-related pre- and post-big bang cosmological phases. 

The string perturbative vacuum is, in general, a higher-dimensional 
state, and the initial growth of the dilatonic coupling $g$ requires, 
according to the lowest-order action, a large enough number of expanding 
dimensions. For instance, in a Bianchi-type I background with $d$ 
expanding and $n$ contracting isotropic spatial dimensions, the growth 
of $g$ requires \cite{6} 
$d+\sqrt{d+n}>n$, which cannot be satisfied by $d=3$, in 
particular, in the ten-dimensional superstring vacuum. With a monotonic 
evolution of the scale factor, this represents another obstruction to 
a smooth transition to our present, 
dimensionally reduced Universe.

The aim of this paper is to show that the initial perturbative vacuum 
is not inconsistent, at the quantum level, with a final contracting
cosmological configuration, 
when we add to the lowest-order action  an 
appropriate dilaton potential (such as the simple one induced by an 
effective cosmological constant). In particular, 
for a WDW potential which is 
translationally invariant in minisuperspace, along the direction 
parametrized by the scale factor, and for which the sign of the Hubble 
factor is classically conserved during the whole evolution,   
the cosmological contraction corresponds to a pure quantum effect. It can be 
described as an ``anti-tunnelling" of the WDW wave function  
from the string perturbative vacuum or, in a 
third quantization \cite{10a} language, 
as a production of ``pairs of universes" 
(one expanding, the other contracting) out of the third quantized 
vacuum. Such a process requires 
the identification of the time-like coordinate in minisuperspace with 
the direction parametrized by the shifted dilaton $\fb$ (see below), and 
is complementary to the process of spatial reflection of the wave 
function, which describes transitions from pre- to post-big bang 
configurations \cite{7}. 

We shall adopt, in this paper, the 
minisuperspace model 
already discussed in \cite{7}, based on the tree-level,
lowest-order in $\ap$, string effective action \cite{11}. Working in the
simplifying assumption that only the metric and the dilaton contribute
non-trivially to the background, in $d$ isotropic spatial dimensions, the 
corresponding action can be written as
\beq
S = -\frac{1}{2\,\lambda_s^{d-1}}\,\int\,d^{d+1}x\,\sqrt{|g|}\,e^{-\phi}
\,\left(R+\partial_{\mu}\phi\partial^{\mu}\phi +V \right ).
\label{21}
\eeq
Here $\lambda_s=(\ap)^{1/2}$ is the
fundamental string length parameter governing the 
higher-derivative expansion of the action, and $V$ is a (possibly
non-perturbative) dilaton potential.  By using the parametrization
appropriate to an isotropic, spatially flat cosmological background:
\beq
g_{\mu\nu} ={\rm diag} \left(N^2(t), -a^2(t) \da_{ij}\right), ~~~~~~~~
a= \exp\left[\b (t)/\sqrt{d}\right], ~~~~~~~~ \phi=\phi(t),
\label{22}
\eeq
and assuming spatial sections of finite volume, the action can be
expressed in the convenient form
\beq
S=\frac{\lambda_s}{2}\,\int\,dt\,{e^{-\fb}\over N}\,
\left(\dot{\beta}^2-\dot{\fb}^2-
N\,V \right)\,,
\label{23}
\eeq
where $\fb$ is the shifted dilaton: 
\beq
\fb=\phi-\log\,\int\,d^dx/\lambda_s^d -\sqrt{d}\,\beta \,.
\label{24}
\eeq
The variation with respect to $N$ then leads to the Hamiltonian
constraint
\beq
\Pi^2_{\beta}-\Pi^2_{\fb}
+\lambda_s^2\,V(\b,\fb)\,e^{-2\,\fb}=0~,
\label{25}
\eeq
where $\pbe,\pfb$ are the (dimensionless) canonical momenta (in the
gauge $N=1$):
\beq
\Pi_{\beta}={\da S\over \da\dot{\beta}}=
\lambda_s\,\dot{\beta}\,e^{-\fb} , ~~~~~~~~~~~~
\Pi_{\fb}={\da S\over \da\dot{\fb}}=
-\lambda_s\,\dot{\fb}\,e^{-\fb} .
\label{26}
\eeq

When $V=0$, the classical solutions of the action
(\ref{23})  describing the phase of accelerated pre-big bang 
evolution are characterized by two duality-related branches \cite{6},
defined in the negative time range: 
\beq
t<0, ~~~~a=a_0(-t)^{\mp 1/\sqrt d}, ~~~~\fb-\phi_0=-\ln(-t)=\pm \b,
~~~~\pbe=\pm k={\rm const}, ~~~~\pfb=\mp \pbe <0
\label{27}
\eeq
($k$, $a_0$ and $\phi_0$ are integration constants). 
For the upper-sign branch the metric is expanding ($\pbe>0$), and the 
curvature scale $\dot\b ^2$ and the string coupling $g(t)$ are 
growing, starting asymptotically from the perturbative vacuum, 
the state with flat metric 
($\bp=0=\fbp$) and vanishing coupling constant ($\phi=-\infty$,
$g=0$). The lower-sign branch corresponds instead to a contracting 
configuration ($\pbe<0$), in which the coupling $g(t)$ is decreasing. 
In the presence of a constant dilaton potential,
$V=\La= {\rm const}$, the accelerated pre-big bang solutions 
are again characterized by two branches \cite{12}:  
\beq
t<0, ~~~~a=a_0\left[\tanh(-t\sqrt{\Lambda}/{2})
\right]^{\mp 1/\sqrt{d}}, ~~~~\fb-\phi_0=-\ln \sinh \left(-t\sqrt{\Lambda}
\right), ~~~~\pfb <0,
\label{29}
\eeq
which are respectively expanding with growing dilaton 
(upper sign, $\pbe>0$) and contracting with decreasing dilaton
(lower sign, $\pbe<0$). In this case 
both branches are 
dominated, in the low-curvature regime,  by the contribution
of a positive cosmological constant $\La$. The initial 
perturbative vacuum is replaced by a
configuration with flat metric and linearly evolving dilaton ($\bp=0$,
$\dot\phi={\rm const}$), another well-known string theory
background \cite{13} (exact solution to all orders
in the $\ap$ expansion). Near the singularity ($t\ra 0_-$), however, the 
contribution of $\La$ becomes negligible, and the solution 
(\ref{29}) asymptotically approaches that of eq. (\ref{27}).

In this paper we shall assume that an effective cosmological constant 
$\La$ is 
generated non-perturbatively in the strong coupling, Planckian regime, 
and we shall use the WDW equation to discuss the possibility of 
transitions, 
induced by $\La$, from the perturbative vacuum to a final configuration 
with contracting metric and decreasing dilaton. We shall consider, in 
particular, 
the case in which the effective dilaton potential can be approximated  
by  the Heaviside step function $\theta$ as 
$V(\b, \fb)= \La ~\theta (\fb)$. The corresponding WDW equation, in the 
minisuperspace spanned by $\b$ and $\fb$, is obtained from the 
Hamiltonian constraint (\ref{25}) through the differential 
representation $\Pi= -i \nabla$:
\beq
\left [ \partial^2_{\fb} - 
\partial^2_{ \beta}
+\lambda_s^2\,\Lambda\,\theta(\fb)\,e^{-2\fb} \right ]\, \Psi= 0 \,.
\label{31}
\eeq
The momentum along the $\b$ axis is conserved,
\beq
[\pbe ,H] =0, ~~~~~~~~~~~~\pbe = \ls \dot{\b} e^{-\fb} = k =
{\rm const} ,
\label{212a}
\eeq
and the general solution of the WDW equation can be factorized as
$\Psi_k(\fb,\b) = \psi_k(\fb) e^{ik\b}$. 

Note that we have assumed a potential $V$ depending explicitly only on 
$\fb$ because the classical evolution of the 
scale factor, in that case, is monotonic, and no contracting 
configuration can be eventually obtained, classically, 
if we start from the isotropic perturbative vacuum.
From a quantum-mechanic point of view, however, the situation is 
different. Indeed, if we assign to $\fb$ the role of time-like 
coordinate, eq. (\ref{31}) is formally equivalent to a 
Klein--Gordon equation with time-dependent mass term. The solution
$\psi_k$ is a linear combination of plane waves for $\fb <0$,
and of Bessel functions \cite{14} $J_{\pm\nu}(z)$, of imaginary index
$\nu=ik$ and argument $z=\la_s\sqrt\La e^{-\fb}$, for $\fb>0$. In
particular, the functions 
\bea
\Psi^{(\pm)}_{k}=\frac{e^{ik\b}}{\sqrt{4\pi k}}\,e^{\mp ik\fb}\,,
~~~~~~~~~~~~~~\fb&<&0 , \\
\Psi^{(\pm)}_{k}=\frac{e^{ik\b}}{\sqrt{4\pi k}}
\,\left(\frac{z_0}{2}\right)^{\mp \nu}\,\Gamma(1\pm \nu)\,
J_{\pm \nu}(z)\,,
~~~~~~~~~~~~~~\fb&>&0 ,
\label{33}
\eea
where $z_0=\la_s\sqrt\La$ and $\Ga$ is the Euler function, 
provide  orthonormal sets of solutions with respect to the 
Klein-Gordon scalar product 
\beq
\label{scal}
(\Psi^1,\Psi^2)=-i\,\int d \beta \, 
\Psi^1(\beta,\fb)\stackrel{\leftrightarrow}{\partial}_{\fb}
{\Psi^2}^*(\beta,\fb)\,\,. 
\label{34}
\eeq

We shall fix the boundary conditions by imposing that, for $\fb<0$, the
Universe is represented by the wave function
\beq
\Psi_{Ik}(\b, \fb<0)=\frac{1}{\sqrt{4\pi k}}\,e^{ ik(\b-\fb )}, 
\label{35}
\eeq
corresponding to a state of growing dilaton  and 
accelerated pre-big bang expansion from the perturbative vacuum, with
$\pbe=-\pfb=k>0$ according to eq. (\ref{27}). The eigenvalue $k$ of $\pbe$
parametrizes the initial state in the space of all classical configurations
(\ref{27}). For $\fb>0$ the wave function is uniquely determined by
the matching conditions for $\Psi$ and $\partial_{\fb} \Psi$
at $\fb=0$, in terms of the functions (\ref{33}), as 
\beq
\Psi_{IIk}(\b, \fb>0)=   A_k^+\,\Psi_{k}^{(+)} +  
 A_k^-\,\Psi_{k}^{(-)} ,
\label{36}
\eeq
where
\beq
A^{\pm}_{k}=\frac{i\,z_0}{2\,k}\,\left
(\frac{z_0}{2}\right)^{\pm ik}\,\Gamma(1\mp ik)
\,\left[\pm{J}_{\mp ik}^{\prime}(z_0)\mp 
\frac{i\,k}{z_0}\,{J}_{\mp ik}(z_0)\right]
\label{37}
\eeq
(a prime denotes differentiation of the Bessel functions with respect to
their argument). 
Given a pure initial state $\Psi_{I}^{(+)}$ of ``positive frequency" $k$,  
the final state is thus a 
mixture of ``positive" and ``negative" frequency modes,
$\Psi_{II}^{(+)}$ and $\Psi_{II}^{(-)}$, satisfying asymptotically the
conditions 
\bea
&&\lim_{\fb \ra \infty} \Psi_{II}^{(\pm)}(\b,\fb) =
\Psi_{\infty}^{(\pm)}(\b,\fb)\sim e^{ik(\b \mp \fb)} , 
\nonumber\\
&&\pbe\Psi_{\infty}^{(\pm)}= -i \pa_\b\Psi_{\infty}^{(\pm)}=
k\Psi_{\infty}^{(\pm)},~~~~~ ~
\pfb\Psi_{\infty}^{(\pm)}= -i \pa_{\fb}\Psi_{\infty}^{(\pm)}=
\mp \pbe\Psi_{\infty}^{(\pm)} .
\label{213}
\eea

The mixing is determined by the 
coefficients $A^{\pm}_{k}$, satisfying the standard Bogoliubov 
normalization condition
$|A_k^+|^2-|A_{k}^-|^2=1$. In a second quantization context, it is well 
known that such a mixing describes a process of pair production 
\cite{15}, the negative energy mode being associated to an 
antiparticle state of positive energy and opposite spatial momentum. It 
thus seems correct to interpret the above splitting of the WDW wave 
function, in a third quantization context \cite{10a}, as the production 
of a pair of universes, with quantum numbers $\{\pbe, 
\pfb \}$, corresponding to positive energy ($\pfb<0$) and opposite 
momentum along the spacelike direction $\b$. One of the two universes is 
isotropically expanding ($\pbe>0$), with growing dilaton; the 
``anti-universe" is isotropically contracting ($\pbe<0$), with decreasing 
dilaton. Both configurations evolve towards the curvature singularity of
the classical pre-big bang solution (\ref{29}). However, while the 
growing dilaton state corresponds to a continuous classical evolution 
from the perturbative vacuum, no smooth connection to such vacuum is 
possible, classically, for the state with decreasing dilaton.

It is important to stress that, as long as $V=V(\fb)$ and, 
consequently, $\pbe$ is 
conserved, a third-quantized production of universes is 
only possible provided we assign the role of time-like coordinate to
$\fb$, and the potential satisfies $V(\fb)e^{-2\fb}$ $\ra 0$ for $\fb \ra 
+\infty$ (in order to identify, asymptotically, positive and negative 
frequency modes). The pairs of universes are produced in the limit of 
large positive $\fb$, so that we cannot describe in this context a 
transition to post-big bang cosmological configurations, which are 
instead characterized by $\fb<0$. A quantum description of the 
transition from pre- to post-big bang requires in fact the 
interpretation of $\b$ as the time-like axis, as discussed in \cite{7}. In 
that case, a third quantized production of pairs becomes possible only 
if $\pbe$ is not conserved, namely if $V$ depends also on $\b$. 

For the process considered in this paper, the probability 
is controlled by
$|A_{k}^-|^2$,  which determines the expectation number pairs of universes
produced in the final state. The production probability is negligible when
$|A_{k}^-|\ll 1$;  it has the typical probability of a vacuum fluctuation
effect when  $|A_{k}^-|\sim |A_{k}^+|\sim 1$; finally, when 
$|A_{k}^-|\sim |A_{k}^+|\gg1$, the initial wave function is parametrically
amplified \cite{15a} and the probability is large.  
In our case, the interesting parameter characterizing the
process, besides $\La$, is the portion of proper spatial volume 
$\Om=a^d\int d^dx$ undergoing the transition. Considering, in 
particular, $d=3$ spatial dimensions, and using the 
definitions of $k$ and $\fb$, the initial momentum $k$ can be
conveniently expressed as $k=\sqrt 3 \Om_s g_s^{-2}\la_s^{-3}$, 
where $g_s=\exp(\phi_s/2)$ and $\Om_s$ are, respectively, 
the value of the coupling and of the proper spatial 
volume evaluated at the string scale $t=t_s$, 
when  $H\equiv \bp/\sqrt 3=\la_s^{-1}$. 
By exploiting the properties of the Bessel functions, we can then 
express the asymptotic limits of
the Bogoliubov coefficients (\ref{37}) in terms of the physical
parameters $\Om_s$ and $\La$. We obtain, at fixed 
$\Om_s/(g_s^2\la_s^3)=1$,
\bea
&&|A^+|^2-1 \simeq |A^-|^2 \simeq  \frac{1}{48}\,\Lambda^{2}\,\lambda_s^4\,
, ~~~~~~~~~~~~~~~~~~~~~~~~~~~~~~~~~~~~~~\La\ll \la_s^{-2},
\label{310}\\
&&|A^+|^2 \simeq |A^-|^2 \simeq  
\sqrt{\Lambda\,\lambda_s^2}\,
\,\frac{\cosh(\sqrt{3}\,\pi) - \sin(2\,\sqrt{\Lambda\,\lambda_s^2})}
{4\,\sqrt{3}\,\sinh(\sqrt{3}\,\pi)}\,, 
~~~~~~~\La\gg \la_s^{-2},
\label{311}
\eea
and, at fixed $\La \la_s^2=1$,
\beq
|A^+|^2 \simeq |A^-|^2 \simeq  
\frac{g_s^4\,\lambda_s^6}{12\,\Omega_s^2}\,|J^{\prime}_0(1)|^2\,,
 ~~~~~~~ |J^{\prime}_0(1)| \simeq 0.44, ~~~~~~~
\Om_s\ll g_s^2 \la_s^3 
\label{312}
\eeq
(the limit $\Om_s\gg g_s^2 \la_s^3$ cannot be performed because the 
quantum process is confined to the region of large $\fb$).

The quantum production of universes in a state with non-vanishing 
cosmological constant $\La$ is thus strongly suppressed 
for small values of $\La$, 
while it is favoured in the opposite limit 
of large $\La$ and of proper volumes that are 
small in string units, in qualitative 
agreement with previous results \cite{7},  and 
also with the general approach to quantum cosmology 
based on tunnelling boundary 
conditions \cite{4,5}. Instead of a ``tunnelling from nothing", however, 
this quantum production of expanding and contracting universes can be 
seen as an ``anti-tunnelling from the string perturbative vacuum" of 
the WDW wave function. Indeed, the asymptotic expansion of the solution 
(\ref{35}), (\ref{36}),
\bea
\fb \ra +\infty &,& ~~~~~~~~~~~ \psi \sim A_{in} e^{-ik\fb}+
 A_{ref} e^{ik\fb},\nonumber \\
\fb \ra -\infty &,& ~~~~~~~~~~~ \psi \sim  A_{tr} e^{-ik\fb}, 
\label{313}
\eea
describes formally a scattering process along $\fb$, in which the 
expanding universe corresponds to the incident part of the wave function, 
the contracting anti-universe to the reflected part, and the initial 
vacuum to the transmitted part. In the parametric amplification regime 
of eqs. (\ref{311}) and (\ref{312}), where $|A^+|\sim |A^-|\gg 1$, the 
reflection 
coefficient $R=|A_{ref}|^2/|A_{in}|^2$ is approximately $1$, and
the Bogoliubov coefficient $|A^-|$, which controls the probability of 
pair production, 
becomes the inverse of the
tunnelling coefficient $T=|A_{tr}|^2/|A_{in}|^2$:
\beq
|A^-|^2= {|A_{ref}|^2\over|A_{tr}|^2}={R\over T}\simeq {1\over T}.
\label{314}
\eeq

In view of future applications, 
we have also computed numerically the Bogoliubov coefficients $A^{\pm}$ 
by discretizing the 
WDW equation with the explicit method \cite{16}, and using the routine Fast
Fourier Transform \cite{17}. A computer simulation, in which the pair 
production process is graphically represented by the scattering and 
reflection of an initial wave packet, has given results 
in complete agreement with
the analytic computation (\ref{37}). 

In conclusion, we have shown in this paper that it is not impossible, 
in a quantum cosmology context, to nucleate universes 
in a state characterized by isotropic contraction 
and decreasing dilaton. The process can be described as the production 
from the vacuum of  
universe--anti-universe pairs in the strong coupling regime, triggered by 
the presence of an effective cosmological constant. 
When $V=V(\fb)$ and $\pbe$ is conserved, the pair-production 
process requires the 
identification of $\fb$ as time-like coordinate in minisuperspace, 
while the transition from 
pre- to post-big bang configurations requires the complementary choice of 
$\b$ as the time-like axis.

The validity of our 
analysis is limited by the very crude approximation (the step potential) 
adopted to modellize the time-evolution of the non-perturbative dilaton 
potential. Also, an appropriate 
potential should depend on $\phi$ (not on $\fb$ as assumed in
this paper); in that case, however, the transition from expansion to 
contraction may be allowed also classically (in an appropriate 
limit), and is represented in minisuperspace as a reflection \cite{18} 
(instead of an anti-tunnelling) of the wave function. In spite of these 
limitations, the analysis of this paper confirms that the WDW approach 
provides an adequate framework for a consistent formulation
of quantum string cosmology, with 
the boundary conditions uniquely prescribed by the choice of
the initial perturbative vacuum. 

\vskip 2 cm 

\section*{Acknowledgements}
We are grateful to Vittorio de Alfaro and Roberto Ricci for 
discussions and clarifying comments. Special thanks are due to Gabriele 
Veneziano for a careful reading of the manuscript and for helpful 
suggestions.

\end{document}